# Ferromagnetism in Fe-doped SnO$_2$ thin films

J.M.D. Coey, A.P. Douvalis, C.B. Fitzgerald, M. Venkatesan

Physics Department, Trinity College, Dublin 2, Ireland

*Abstract:* Thin films grown by pulsed-laser deposition from targets of Sn$_{0.95}$Fe$_{0.05}$O$_2$ are transparent ferromagnets with Curie temperature and spontaneous magnetization of 610 K and 2.2 Am$^2$kg$^{-1}$, respectively. The $^{57}$Fe Mössbauer spectra show the iron is all high-spin Fe$^{3+}$ but the films are magnetically inhomogeneous on an atomic scale, with only 23 % of the iron ordering magnetically. The net ferromagnetic moment per ordered iron ion, 1.8 $\mu_B$, is greater than for any simple iron oxide. Ferromagnetic coupling of ferric ions via an electron trapped in a bridging oxygen vacancy (F center) is proposed to explain the high Curie temperature.

First generation spin electronics[1] was based on magnetoresistive sensors and memory elements using electrodes made from alloys of the ferromagnetic 3d metals Fe, Co and Ni. There is an ongoing quest for ferromagnetic semiconductors with a Curie temperature well above room temperature, which could be used for a second generation of spin electronics, as well as a search for transparent ferromagnets which can add an optoelectronic dimension. Much recent interest has been generated by high temperature ferromagnetism in oxide and nitride materials such as ZnO with Co or Mn doping[2-4], $TiO_2$ (anatase) with Co[5], GaN with Mn[6], AlN with Cr[7] and $SnO_2$ with Mn[8] or Co[9].

Doubts linger as to whether these are homogeneous, single-phase materials, particularly since the well-accepted mechanisms for ferromagnetic coupling – via spin-polarized p-band holes as in $Ga_{1-x}Mn_xAs$[10], or via double exchange as in mixed valence manganites[11] – do not seem to apply in these oxides and nitrides.

Following a recent report by Ogale et al[9] of high-temperature ferromagnetism with a giant cobalt moment in Co-doped $SnO_2$, we undertook an investigation of the magnetism of Fe-doped $SnO_2$. We find atomic-scale inhomogeneity and remarkably strong ferromagnetism, for which a novel ferromagnetic exchange mechanism is suggested.

Ceramic targets of $Sn_{0.95}Fe_{0.05}O_2$ were first prepared by solid-state reaction of $SnO_2$ and FeO or $^{57}Fe_2O_3$ at 1150 °C. Rietveld analysis of the X-ray diffraction patterns of the targets showed $SnO_2$ with a trace of $\alpha$-$Fe_2O_3$ (Fig. 1). Elemental maps of the targets obtained by energy-dispersive X-ray diffraction (EDAX) indicated a non-uniform iron distribution, with some tendency to accumulate iron in regions 2- 4 μm in size which were identified as Sn-doped hematite. The ceramics were ferromagnetic, with a magnetization at 5 K of 2.3 Am$^2$kg$^{-1}$ ($\approx$ 1.2 $\mu_B$/Fe), and a Curie temperature of 360 K. Mössbauer spectra showed that all the iron was high-spin Fe$^{3+}$, and 88 % of it was magnetically ordered with a hyperfine field of 53.3 T at 19 K. The ferromagnetism cannot be attributed to the Sn-doped hematite, which is a canted antiferromagnet with a weak net moment[12].

The thin films were deposited on R-cut sapphire substrates using a KrF excimer laser operating at 248 nm and 10 Hz. Laser fluence on the target was 1.8 Jcm$^{-2}$. The target-substrate distance was 35 mm and the oxygen pressure was 10$^{-4}$ mbar. Unlike the targets, the films show no sign of any micron-scale nonuniformity in the distribution of iron or tin in EDAX analysis. The X-ray diffraction patterns (Fig. 1) show the rutile-structure films to be single-phase and well-oriented with a (101) texture. Film thickness was monitored during deposition using optical reflectivity at 635 nm, and it was independently calibrated by direct measurement in an atomic force microscope of a trench milled by Argon-ion etching. Thicknesses were in the range 100 - 300 nm. Root mean square surface roughness determined by atomic force microscopy (AFM) was generally less than 5 nm. Films showed a 'blister-like' texture, with relief on a lateral scale of approximately 200 nm. Magnetic contrast found in magnetic force microscopy (MFM) appeared to decorate the topology, consistent with uniform in-plane magnetization of the films. The films were transparent, with a faint brown tinge. Optical absorption spectra showed a red shift of the absorption edge from 330 nm for undoped SnO$_2$ to 350 nm in the iron-doped films. After heating in vacuum at 780 K, the films behave as narrow gap semiconductors with an activation energy of 75 meV at room temperature.

Magnetization measurements were made in a SQUID magnetometer with a high temperature furnace. The films are ferromagnetic at room-temperature, with moments ranging from 2 – 9 Am$^2$ kg$^{-1}$ and evident hysteresis (Fig. 2). The larger moments were in films made with FeO, but the analysis here is based on the $^{57}$Fe sample. The Curie temperature obtained from a thermomagnetic scan in a field of 150 mT was 610 ± 10 K. Analysis of the homogeneity of the magnetization on an atomic scale is possible from the room temperature Mössbauer spectrum of a film made from the $^{57}$Fe target, shown in Fig 3. The film thickness was 190 nm, and the Fe:(Fe + Sn) ratio was 14%. Invariably the ratio of transition-metal to tin in films was very much greater than that in the targets. The Mössbauer spectrum consists of two main components, both due to high-spin ferric iron, with isomer shifts of 0.29 and 0.38 mm s$^{-1}$ relative to iron metal. One is the central paramagnetic doublet, with quadrupole splitting 0.81 mm s$^{-1}$, which accounts for 77 % of the absorption area. This iron is not magnetically ordered. The other component is a magnetic sextet with a hyperfine field of 51.2 T, which accounts for the remaining 23 % of the absorption. The film is therefore magnetically inhomogeneous on an atomic scale, with only about one iron ion in four being magnetically ordered. After taking this into account, the measured magnetization corresponds to a net ferromagnetic moment of 1.8 $\mu_B$ per ordered iron. It is clear from the hyperfine field that the ferric moment is 5 $\mu_B$ per ion, so the magnetic order involves some sort of ferrimagnetic compensation such as ↑↑↓..... It is noteworthy that the net iron moment is greater here than in any simple iron oxide; for example the net moment per iron in magnetite (Fe$_3$O$_4$) is 4/3 = 1.3 $\mu_B$, whereas that in YIG (Y$_3$Fe$_5$O$_{12}$) is 5/5 = 1.0 $\mu_B$. The picture of the magnetic order is therefore one of extended ferromagnetic regions with some reversed spins, and many isolated paramagnetic iron sites. The hysteresis, the Mössbauer spectrum and the magnetic force microscopy indicate that the ferromagnetic regions are extended, and not small isolated clusters.

It is difficult to envisage how the observed moment could be achieved from purely antiferromagnetic superexchange in the rutile structure. We are therefore led to consider an underlying ferromagnetic exchange mechanism in these films, which should involve the oxygen vacancies that will arise naturally to ensure charge neutrality when a trivalent ion is substituted in $SnO_2$. We expect that $Fe^{3+}$ – [] – $Fe^{3+}$ groups will be common in the structure, where [] denotes an oxygen vacancy. An electron trapped in the oxygen vacancy constitutes an F-center, where the electron occupies an orbital which overlaps with the d-shells of both iron neighbours. The radius of the electron orbital is of the order $a_o\varepsilon$, where $a_o$ is the Bohr radius and $\varepsilon$ is the dielectric constant of $SnO_2$ (~13)[13]. Since $Fe^{3+}$, $3d^5$, only has unoccupied minority spin orbitals, the trapped electron will be ↓ and the two iron neighbours ↑ (Fig. 4).

This direct ferromagnetic coupling we call F-Center Exchange (FCE), where the F-center resembles Kasuya's bound magnetic polaron[14]. We expect FCE to be the dominant exchange mechanism in the $SnO_2$:Fe films. The average ferromagnetic moment per iron is reduced from 5.0 to 4.5 $\mu_B$. Further reduction will result from any antiferromagnetic $Fe^{3+}$ – $O^{2-}$ – $Fe^{3+}$ superexchange bonds.

In summary, we have found that $Sn_{1-x}Fe_xO_2$ is a transparent ferromagnet with an exceptionally large net moment per ordered ferric ion. The Curie temperature is high, although the films are magnetically inhomogeneous at the atomic scale. A new ferromagnetic exchange mechanism is suggested which involves a spin-polarized electron trapped at an oxygen vacancy. Further work on these oxide ferromagnets should focus on increasing the mobility of the spin-polarized F-center electrons, and developing the link with high-k dielectrics.

Acknowledgements:

This work was supported by Science Foundation Ireland and the HEA. We are grateful to Oscar Céspedes Boldoba for measurements of film thickness and MFM, and Prof. James Lunney for use of his laboratory facilities.

Fig 1 X-ray diffraction patterns of an $Sn_{0.95}Fe_{0.05}O_2$ ceramic target, and a corresponding thin film deposited by pulsed-laser deposition. The film is well oriented with a rutile (101) texture: "S" indicates substrate peaks.

Fig 2 Typical magnetization curve of an $Sn_{0.86}Fe_{0.14}O_2$ thin film at room temperature. Inset shows the temperature dependence of magnetization indicating the Curie temperature transition.

Fig 3 Mössbauer spectrum of a $Sn_{0.86}Fe_{0.14}O_2$ thin film at room temperature, showing a central paramagnetic doublet and a hyperfine-split pattern due to magnetically-ordered iron. All iron is high-spin $Fe^{3+}$.

Fig 4 Schematic comparison of (a) superexchange (SE) and (b) F-center exchange (FCE)

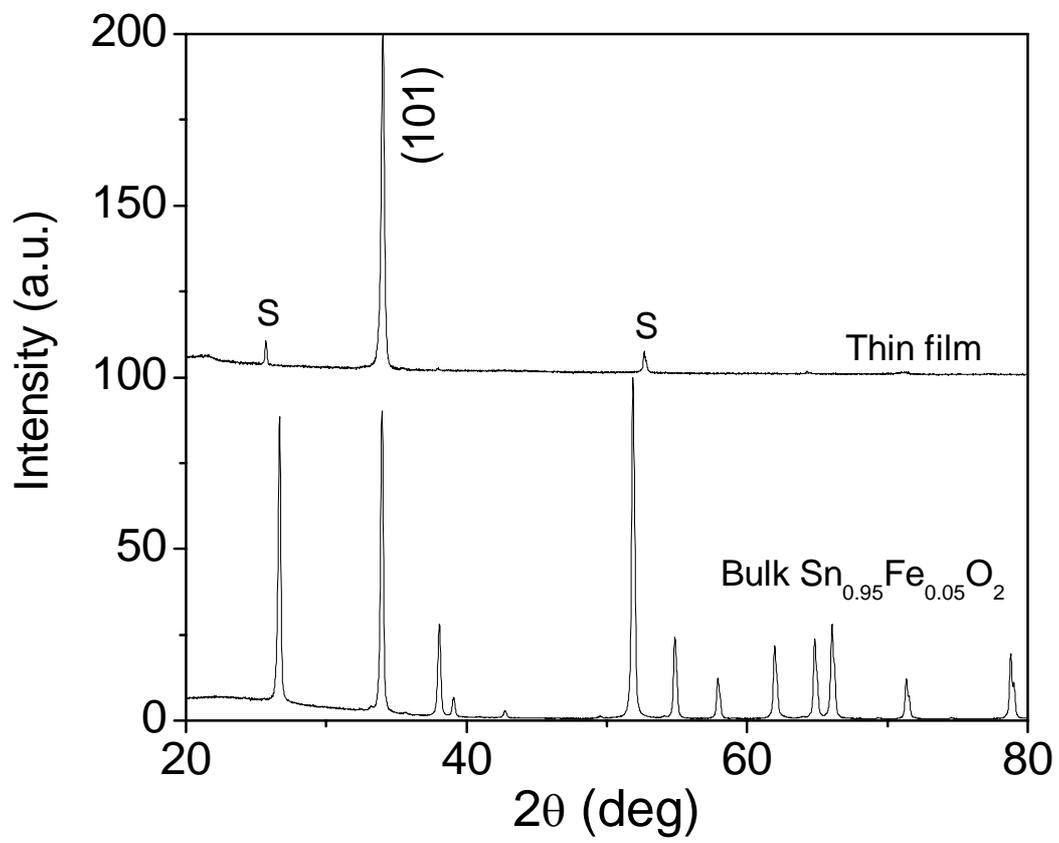

Figure 1
J. M. D. Coey et al

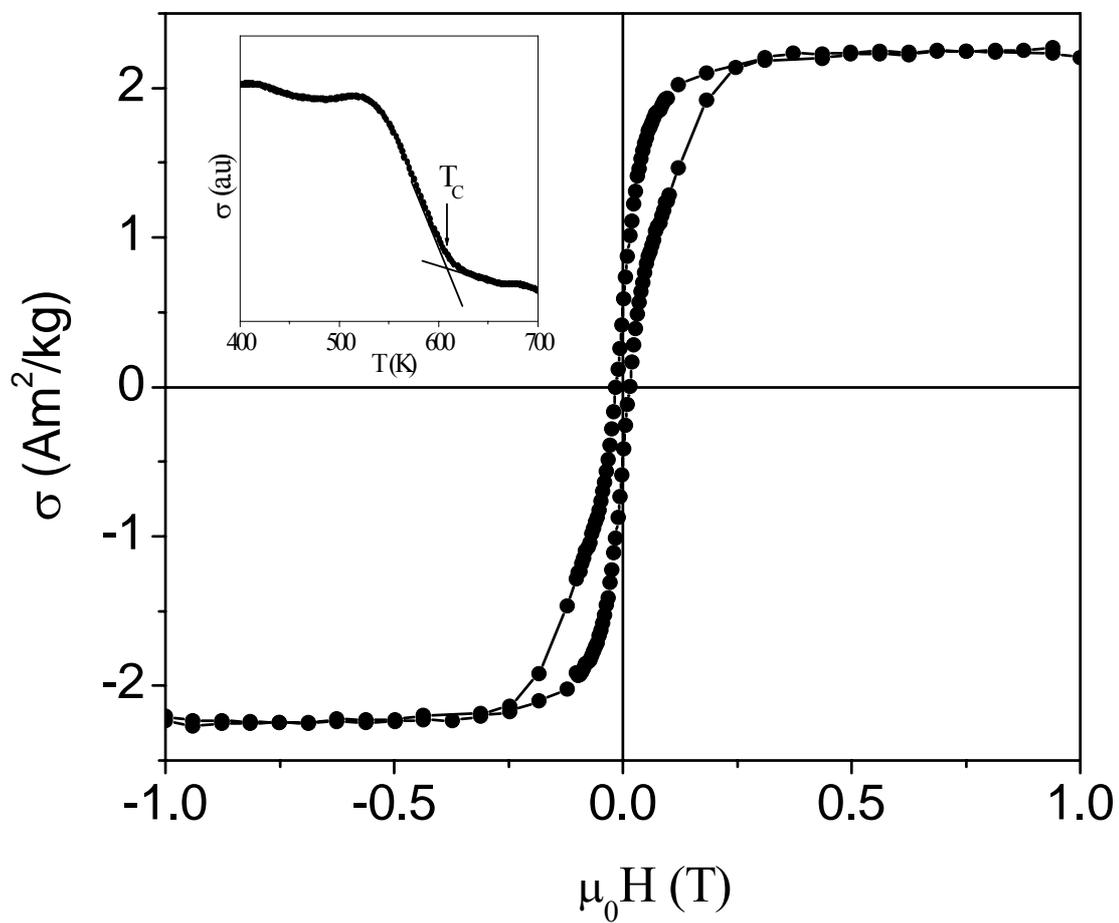

Figure 2
J. M. D. Coey et al

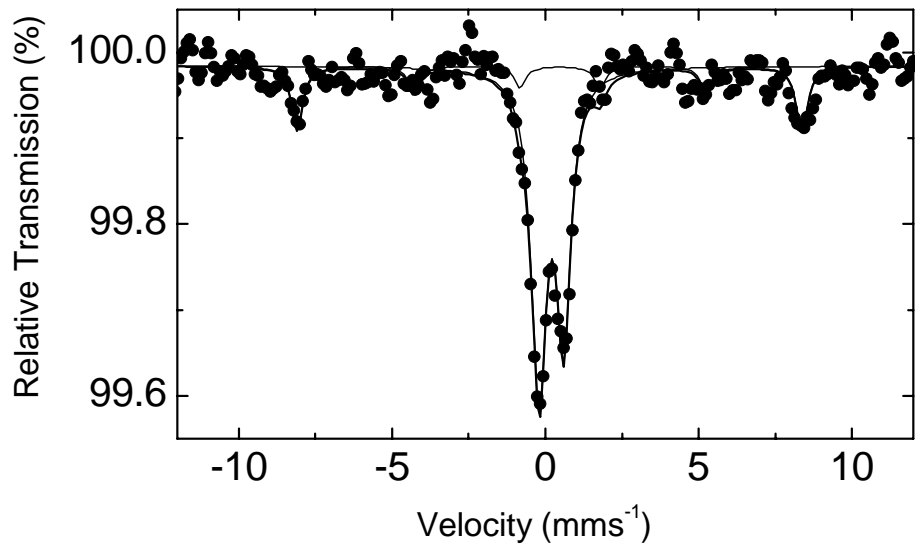

Figure 3
J. M. D. Coey et al

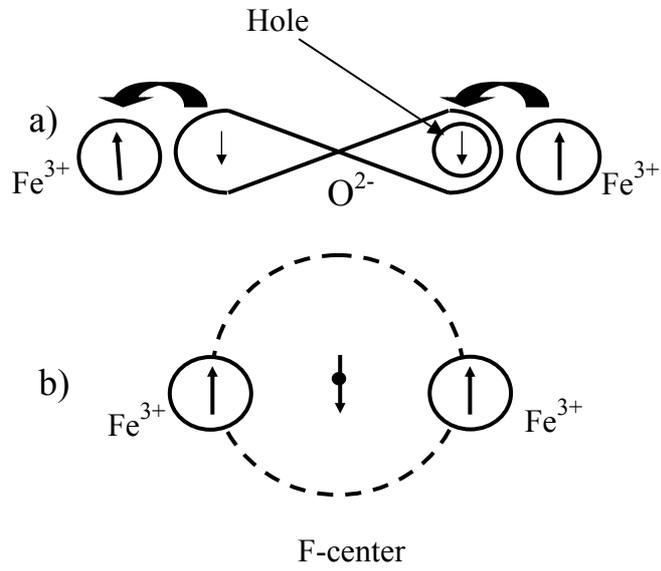

Figure 4
J. M. D. Coey et al